Revised version 5 June 2017

# Measuring the radius and mass of Planet Nine

J. Schneider

Observatoire de Paris, LUTh-CNRS, UMR 8102, 92190, Meudon, France
*jean.schneider@obspm.fr*

**Abstract**

Batygin and Brown (2016) have suggested the existence of a new Solar System planet supposed to be responsible for the perturbation of eccentric orbits of small outer bodies. The main challenge is now to detect and characterize this putative body. Here we investigate the principles of the determination of its physical parameters, mainly its mass and radius. For that purpose we concentrate on two methods, stellar occultations and gravitational microlensing effects (amplification, deflection and time delay). We estimate the main characteristics of a possible occultation or gravitational effects: flux variation of a background star, duration and probability of occurence. We investigate also additional benefits of direct imaging and of an occultation.

**keywords** Planet P9 - stellar occultations – gravitational lensing

**1. Introduction**

Batygin and Brown (2016) have suggested the existence of a new planet in the outer Solar System (called hereafter P9), supposed to be responsible for the perturbation of eccentric orbits of small bodies. From the characteristics of the observed perturbations of orbits, its mass M9 is estimated to be around 10-30 $M_\oplus$ and its distance is estimated to be around 700 AU (Brown & Batygin 2016). More recent estipmate derived the the Cassini data analyis give rather a distance of 1000 AU . For the numerical estimates below, we take the new distance value. The main challenge is now to detect this putative body and, if it exists, to measure its physical characteristics. Several authors have investigated its search by direct imaging. This search presents three issues: constrain the region on the sky plane where the putative planet is localized, estimate its brightness and choose the best possible imaging facilities and search strategies to detect the object. The success of this search depends on the brightness of the object. For a given distance the brightness depends on the planet albedo and radius. The estimates of the brightness run from V = 19.9 to V = 22.7 (Linder & Mordasini 2016) [1].

The advantage of the direct imaging search is that it can be performed any time along the orbit and would be the first step to take spectra of the object. But it cannot provide the size and mass of the planet. Without a value for the radius one cannot infer the planet albedo from its brightness, while this parameter is essential to constrain physical characteristics of the planet surface. A value of the mass would constrain the planet internal structure, its formation scenarii and the mechanism of perturbation of small bodies orbits. Its predicted angular diameter (depending on the exact distance and size), being at most a few tens mas, only an interferometer with a baseline of the order of a kilometer could measure this radius at 1 micron with a 3% precision, a presently difficult task for a V > 20 object. That is why we propose to search for stellar occultations by P9 to measure its radius. For the mass determination, we investigate further the microlensing by P9 of a background star, as already proposed by Philippoc and Chobanu (2016).

**2. Stellar occultations by P9**

In this section we estimate the parameters of occultations and microlensing of background stars by P9.

*2.1 Occultation depth*

Assuming a mass of 30 terrestrial mass and a standard rocky composition, the planet radius R9 is, according to Linder & Mordasini (2016), 2.9 to 6.3 $R_\oplus$, depending on its evolution track. At a distance of 1000 AU, the angular diameter $\delta 9$ is 40 mas for a 3 $R_\oplus$ planet. The angular diameter of the potentially eclipsed background stars are all less than 57 mas (for R Doradus). Therefore, neglecting grazing occultations, during an occultation the star will completely disappear behind P9 and its flux will vanish. More exactly, the star+P9 total flux will drop to the P9 flux, which is of the same order of magnitude, or higher, for faint stars.

For a star with stellar flux $F_*$ fainter than P9, the photometric accuracy $\Delta F9$ of the P9 flux required to detect an occultation with and SNR of 3 must be better than $F_*/3$.

---

[1] We have renormalized to a 1000 AU distance the Linder \& Mordasini (2016) estimate
of V = 20.6 to V = 23.2 for a distance of 700 AU.

*2.2 Occultation duration*

The P9 apparent motion is a composition of its parallax $\pi_9$ and its orbital motion. For a 1000 AU distance of P9, its orbital velocity $v_9$ is about 1 km/sec, giving an apparent motion of 70 arcsec/yr on the sky plane. The latter dominates the proper motion of bakground stars, being all less than 10 arcsec/yr (for Barnard's star). The Earth orbital velocity, 30 km/sec, being much larger than the P9 orbital velocity, it dominates the apparent motion (in the Earth frame) of the observed occultation shadow. If the star path lies on the equator of the planet, the occultation duration is then

$$D_{occ} = 2R_9/(30\,km/sec)/\cos\alpha_\oplus$$

where $\alpha_\oplus$ is the angle between the P9 line of sight and the Earth orbital velocity vector at the time of occultation.

*2.3 Occultation geometry and kinematics*

The above estimate of $D_{occ}$ assumes implicitly an occultation of the background star at the planet equator. Since the true impact parameter of the occultation, the distance between the occultation path and the planet center, is not known, one cannot *a priori* infer the planet radius from the relation

$$2R9 = 30\,D_{occ}\cos\alpha_\oplus \text{ km/sec}$$

This problem occurs for any stellar occultation by Solar System bodies, but then it is generally solved by the accumulation of several occultations by the same object. For instance, for the minor planet Ixion, up to 425 stellar occultations are predicted (Assafin et al. 2012). Since in our case occultations are rare (see the section "Occultation probability"), one cannot use this approach. We thus now show that the measurement of the duration of the ingess and egress of the occultation can solve the problem.

The occultation duration $D_{occ}$ depends on the geometry of the occultation and on the kinematics of P9 and of the occulted star. The geometry of the occultation is represented on Figure 1.

The paths of the planet and of the star are generally not parallel. Consequently, as shown on Figure 1, the impact parameter, i.e. the distance between the star path and the planet center, are different at ingress and egress. We designate their angular sizes by $\delta_{b_{in}}$ and $\delta_{b_{out}}$.

The angular size of the occultation path ("chord") is given by

$$\delta_{occ} = \delta_1 + \delta_2 = \sqrt{\delta_9^2 - \delta_{b_{in}}^2} + \sqrt{\delta_9^2 - \delta_{b_{out}}^2} \qquad (1)$$

where the different parameters of equation (1) are represented on Figure 1. We assume that during the ingress and the egress the impact parameter is approximately constant. Therefore the angular sizes of the ingress and egress pathes are given by

$$\delta_{in} = \sqrt{(\delta_9 + \delta_*)^2 - \delta_{b_{in}}^2} - \sqrt{(\delta_9 - \delta_*)^2 - \delta_{b_{in}}^2} \qquad (2)$$

$$\delta_{out} = \sqrt{(\delta_9 + \delta_*)^2 - \delta_{b_{out}}^2} - \sqrt{(\delta_9 - \delta_*)^2 - \delta_{b_{out}}^2} \qquad (3)$$

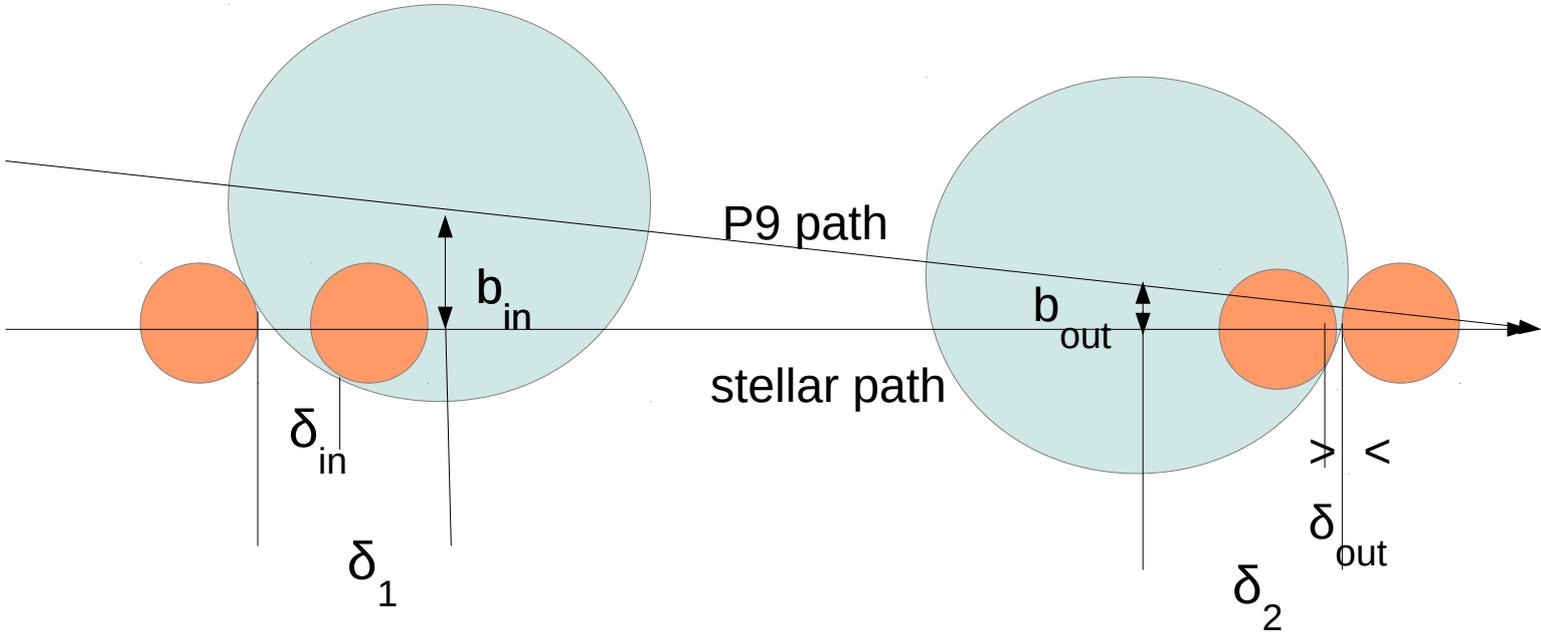

**Figure 1**  Kinematics and geometry of a stellar occultation by P9

From equations (1), (2) and (3) one has

$$\delta_9^2 = \frac{\delta_{occ}^2 (\delta_*^2 \delta_{in}^2 + \delta_*^2 \delta_{out}^2 - 2) + 2\delta_{occ}^2 \sqrt{4\delta_*^4/(\delta_{in}^2 \delta_{out}^2) + 1 - (\delta_*^2/\delta_{in}^2 + \delta_*^2/\delta_{out}^2)}}{\delta_*^2 (1/\delta_{in}^2 - 1/\delta_{out}^2)}$$

The geometrical parameters $\delta_{occ}$, $\delta_{in}$ and $\delta_{out}$ are not primary observables. But they can be derived from their corresponding durations $D_{occ}$, $D_{in}$ and $D_{out}$ which are related to, them by

$$D_{occ} = \delta_{occ}/(\omega_9 - \omega_*) \qquad (4)$$

$$D_{in} = \delta_{in}/(\omega_9 - \omega_*) \qquad (5)$$

and

$$D_{out} = \delta_{out}/(\omega_9 - \omega_*) \qquad (6)$$

where $\omega_9$ is the angular speed of P9, projected onto the stellar path and $\omega_*$ is the angular star speed, i.e. its proper motion $\mu_*$. $\omega_9$ is the composition of the parallactic speed $\omega_{\pi 9}$ and the orbital angular velocity $v_9$ of P9: $\omega_9 = (\omega_{\pi 9} + v_9 \sin i)\cos\beta$, where *i* is the inclination of the P9 orbit relative to the ecliptic plane. $\beta$ is the angle between the stellar and P9 paths on the sky plane. The later angle is inferred from direct images of the P9 and star trajectories on the sky, supposed to be taken in addition to the occultation observations. The orbital inclination *i* of P9 is inferred from the best fit of the perturbations of the Solar System small bodies by P9 (Brown & Batygin 2016).

The angular size $\delta_9$ can finally be derived from observables $D_{occ}$, $\omega_9$, $\omega_*$, $D_{in}$, and $D_{out}$, through equations (4) to (6). The radius R9 itself can be derived from $\delta_9$ and from the P9 parallax $\pi_9$.

**3. Gravitational lensing**

Recently Philippov and Chobanu (2016) have proposed to detect P9 by gravitational deflection and lensing of the background star. Here we develop further numerical estimates and probability of occurence.

*3.1 Gravitational amplification*

For a star at an apparent distance *r* from P9 at the planet distance, the essential parameter to evaluate the amplification is the Einstein radius $R_{E9}$ of P9. It is given by

$$R_{E9} = \sqrt{R_{S9} D9}$$

Here *D9* is the P9 distance and $R_{S9}$ its Scharzschild radius $GM9/c^2$. In the approximation of a point-like stellar source, at a distance *r* from the P9 center, the gravitational amplification $A_G$ is given by (Schneider 1989)

$$A_G(r)=\frac{s^4}{s^4-R_E^4}$$

where $s=(r+\sqrt{r^2+R_{E9}^2})/2$ For a $30\,M_\oplus$ mass and a 1000 AU distance, $R_{S9}=3.3\ 10^8$ cm, while $R_{E9}$ = 20 $10^8$ cm, or about *R9* for a 3 $R_\oplus$ radius. Therefore, a large part of the gravitational amplification phase will be occulted by the planet. Nevertheless, when the star emerges from the planet disc at a distance *r* larger than *R9*, the flux increase is still not negligible:

$$A_G(r)-1 \sim 4R_{E9}/r$$

with a value $A_G$ about 4 at $r=R9$ The flux increase drops down to 10% at a distance $r=r_{10}=40\,R_{E9}$ Note that the gravitational amplification takes place at a detectable level larger than 10% even when the star is not occulted, as long as its line of sight passes at a distance less than about 40 $R_{E9}$ from P9, *i.e.* about 40 *R9*. The corresponding probability of occurence is then 40 times the occultation probability.

## *3.2 Deflection*

The gravitational deflection, given by $\alpha(r)=R_{S9}/r$ for a line of sight at a distance *r* from P9, is 1 arc sec for $r=r_{10}$ for a $M9=30\,M_\oplus$ planet. For a background star at 1 arcsec of P9 at 1000 AU $\alpha$ is 110 µas. Even for very faint stars this kind of deflection will be detectable with large telescopes. Combined with gravitational lensing, the measurement of the deflection would constrain further the planet mass.

## *3.3 Time delay*

A time delay $\Delta T$ for the time of arrival of pulses from a background source has two origins : the « Shapiro » time delay $\Delta T_E$ due to the Einstein effect of retardation of clocks and a geometrical effect $\Delta T_{Geom}$ due to the bending of the light rays passing close to P9.

They are given by

$$\Delta T_E=\frac{2GM9}{c^3}(1+\ln(r_s\frac{D9}{r_s^2}))$$

and

$$\Delta T_{Geom} = \frac{r_S GM9}{2rc^3}$$

where *r* is the impact parameter of the line of sight of a source at a distance $r_S$. They are typically of the order of 10-10 sec. There is no known astrophysical source with pulses sufficiently short to make this kind of time delay measurable. Nevertheless, we present in the conclusion a configuration where these time delays will be measurable, namely by clocks on-bard an *in-situ* mission..

*3.4 Microlensing duration*

Reminding that *b* is the impact parameter of the stellar path (minimum distance of the star to P9 on the sky plane). For a velocity *v* of the apparent position of P9 on the sky plane (due to the Earth orbital motion), the time variation of the amplification $A_G(t)$ is still given by equation (1), where now $s = s(t) = (\sqrt{v^2 t^2 + b^2} + \sqrt{v^2 t^2 + b^2 + R_{E9}^2})/2$

**4. Numerical estimates**

*4.1 Duration of the occultation and ingress/egress*

As we are essentially interested by orders of magnitudes, and since the P9 parallactic motion dominates over its orbital motion and the star's proper motion, we neglect the two latter. Then, for a $R_\odot$ star at 100 pc with an angular diameter $\delta_* $ = 0.5 mas, the ingress/egress duration $D_{in/out}$ is about 12.5 sec for an equatorial occultation.

The occultation duration, depending on the position of the Earth on its orbit at the time of occultaion will be $2R_9/(30\,km/sec)/\cos\alpha_\oplus$ , *i.e.* at least 1300 sec (for $\cos\alpha_\oplus = 1$ ).

*4.2 Probability of occultations*

The *a priori* probability $P_{occ}$ of an occultation by P9 depends on the duration of the survey and the surface density number of the sample of surveyed stars. The later depends on their limiting magnitudes and on the involved sky region. $P_{occ}$ is given by

$$P_{occ} = N_* A_9 \qquad (5)$$

where $N_*$ is the number of stars per arcsec$^2$ and $A_9$ the area of the sky band swept by P9 during the survey. Since the parallactic motion $\delta_{\pi 9}$ dominates over the P9 orbital motion and the stellar proper motions, it is given, at first approximation, by $A_9 = 2\delta_{\pi 9}\delta_9$. For a $3R_\oplus$ planet at 1000 AU, $\delta_9$ is about 40 mas and $\delta_{\pi 9}$ is about 200 arc sec, so that $A_9$ is about 24 arcsec$^2$.

To refine our estimate of $N_*$, it is suitable to have an idea of the current localization of P9 on its orbit. The analysis of the perturbations of Solar System small bodies can only give the orbital elements planet 9. According to Fieng $\omega$ = 150 deg and $\Omega$ = 113 deg. This dynamical analysis cannot give the present position of P9 on its orbit. Nevertheless a recent estimate by Millholland & Laughlin (2017) is RA = 40 ±10 deg and DEC = 0 ±20 deg. In this region, according to Allen (1976), the number of star brighter than V = 21 is $10^4$ per square degree.

The planet spans its sky band $A_9$ in 6 months. It results from equation (5) that the probability of occultation of a V = 21 star in a 6 months survey is about 0.5 $10^{-5}$. This very low probability forces to search for occultations of fainter stars. Applying the rule $N_*(m)/N_*(n) = 2^{m-n}$, to have a 10\% probability it is necessary to monitor V = 32 stars. For a subsequent 6 months survey, the background stars and P9 will have moved and the sky configuration refreshed, so that the probability of an occultation during a second 6 months survey is to be added to the previous one. More generally, for a $N$ times 6 months survey, the occultation probability is 0.005 $N$, *i.e.* about 20% in a 10 year survey of V= 32 stars. Such faint stars will be detectable with the coming 30m-class telescopes.

## 5. Detection strategies

One could *a priori* search "*ab initio*" for occultations and gravitational events (*i.e.* Amplification and deflection), prior to the detection of P9 by direct imaging. But, given that these events are short, with no prediction of occurence, hidden in a large portion of the sky and require large telescopes to achieve the required photometric precision, it is better to search for a direct image of P9 before onje of these events. Indeed, if a direct image is detected somewhere in the sky, it will be easy to extrapolate its apparent trajectory in the sky due to the Earth orbital motion. From there, given an appropriate knowledge of the field of background stars in a band $A_9 = 2\delta_{\pi 9} * \delta_{10}$ (where $\delta_{10} = 40\delta_9$ is the angular distance from P9 where the flux excess due to microlensing exceeds 10%), it will be easy to predict the time of occurences of occultation, microlensing and and gravitational deflection.

**6. Conclusion**

Direct detection of P9 and the detection of stellar occultations and gravitational effects by P9 are complementary. Their combination would allow to infer both the planet radius, mass and its albedo. In the future, once the planet P9 has been detected, if it exists, further observations will be interesting. The detection of a stellar occultation by P9 would have additional benefits.

- It would be possible to investigate a possible atmosphere by spectroscopy of occultation with 30 meter class telescopes.

- It would allow to detect rings (Barnes & Fortney 2004) and satellites (Sartoretti & Schneider 1999), thanks to the shape of the occultation light curve, and possibly surface irregularities ("mountains"), inferred from an asymmetry between the ingress and egress shapes..

High resolution spectroscopy of the planet would allow to determine its radial velocity and thus the eccentricity of its orbit. Indeed, for an eccentricity e, tthe radial velocity as seen from the barycenter of the solar system has amximum given by $VR = e\sqrt{GM_\odot}/\sqrt{(a(1-e^2))}$

For *a* = 1000 AU and an eccentricy 0.6 it is about 600 m/sec detectable in the spectrum of the Sun reflected by P9 with moderate resolution spectrographs connected to large telescopes.

Finally, needless to say, planet P9 would be an exquisite target for future *in situ* missions. One would not be able to explore in deep all the P9 characteristics with ground and space telescopes, but an *in situ* mission would investigate characteristics invisible from Earth, like the composition of its potentially rocky surface and its relief. It would also allow to measure the mass of P9 by the perturbation of the spacecraft trajectory and the measurement of the gravitational time delay of ultra-short on board pulses. And, given its large distance, it would be a useful precursor to consolidate the technology of future interstellar missions.